\documentclass[useAMS,usenatbib,letters]{mn2e}
\usepackage{url,times,graphicx,amsmath,amsfonts,amssymb,aas_macros,color,ulem}

\definecolor{darkgreen}{rgb}{0.0,0.5,0.0}

\newcommand{\beq}{\begin{equation}}
\newcommand{\eeq}{\end{equation}}
\def\beqa{\begin{eqnarray}}
\def\eeqa{\end{eqnarray}}
\def\hmpc{h^{-1}\,{\rm Mpc}}
\def\hkpc{h^{-1}\,{\rm kpc}}
\def\lcdm{\ensuremath{\Lambda}CDM}

\title[Structure formation in scalar-interacting DM]
      {Boosting hierarchical structure formation with scalar-interacting dark
        matter}
\author[Wojciech A. Hellwing, Steffen R. Knollmann and Alexander Knebe]
       {Wojciech A. Hellwing$^{1}$\thanks{E-mail: pchela@camk.edu.pl},
        Steffen R. Knollmann$^{2}$ and Alexander Knebe$^{2}$\\
        $^{1}$Nicolaus Copernicus Astronomical Center, ul. Bartycka 18,
              Warsaw, 00-716, Poland\\
        $^{2}$Grupo de Astrof\'isica, Departamento de F\'isica Te\'orica,
              M\'odulo C-15, Facultad de Ciencias,
              Universidad Aut\'onoma de Madrid,\\
              28049 Cantoblanco, Madrid, Spain}
\begin{document}

\date{Accepted 2010 August 22 . Received 2010 August 11; in original form 2010 April 16}

\pagerange{\pageref{firstpage}--\pageref{lastpage}} \pubyear{2010}

\maketitle

\label{firstpage}

\begin{abstract}
We investigate the effect of long-range scalar interactions in dark
matter (DM) models of cosmic structure formation with a particular focus
on the formation times of haloes. Utilising $N$-body simulations with
$512^3$ DM particles we show that in our models dark matter haloes form
substantially earlier: tracing objects up to redshift $z\sim6$ we find
that the formation time, as characterised by the redshift $z_{1/2}$ at
which the halo has assembled half of its final mass, is gradually
shifted from $z_{1/2}\approx 1.83$ in the fiducial \lcdm\ model to
$z_{1/2}\approx 2.54$ in the most extreme scalar-interaction model. This
is accompanied by a shift of the redshift that marks the transition
between merger and steady accretion epochs from $z_{*}\approx 4.32$ in
the \lcdm\ halos to $z_{*}\approx 6.39$ in our strongest
 interaction model. In other words, the scalar-interacting model
 employed in this work
produces more structures at high redshifts, prolonging at the same time
 the steady accretion phases. These effects
taken together can help the \lcdm\ model to account for a high
redshift reionisation as indicated by the WMAP data and can alleviate
issues related to the survival of the thin-disk dominated galaxies at
low redshifts.
\end{abstract}


\section{Introduction}

The $\Lambda$ cold dark matter (\lcdm) model has proven to be capable
of explaining a tremendous amount of observational data. As we entered
the era of precision cosmology --- both observationally and from a
modellers perspective --- we are left with a more and more detailed
picture of what actually took place when the Universe cooled down and
formed structures starting from $z\sim1100$ to the present day. But
there are still issues related to the structure formation on galactic
scales to be resolved. For instance, observations of high redshift
quasars \citep{quazars_z6_reionisation} and galaxies
\citep{early_reionisation} imply that the epoch of the reionisation of the
Universe ended at $z\sim 6$.  On the other hand observations of the
polarisation of the cosmic microwave background radiation
\citep{WMAP07,reion2,reion3} suggest that the Universe was already reionised at
$z{\stackrel{>}{{}_\sim}}10-11$.  However, galaxy formation within the
\lcdm\ model using the suggested moderate normalisation of the power
spectrum of primordial density perturbations $\sigma_8=0.8$ \citep[as
favoured by recent WMAP observations; ][]{WMAP07} may not happen early enough
to account for such a high redshift of reionisation.  Furthermore the
astonishing array of recent observations of ultra-faint distant
objects \citep{high_z_evolution_LF,z_10_galaxies} as well as advancements
in the understanding of the high-$z$ physics
\citep{star_formation_z8,supermassive_BH_growth1,supermassive_BH_growth2}
have shown that the young Universe was a busy and already structure-rich
place.  In addition, there are issues related to the survival of
thin-disk dominated galaxies.  It was shown that in a pure CDM model of
galaxy-halo formation, the majority of Milky-Way-sized haloes with
$M_0\sim10^{12}h^{-1}M_{\odot}$ have accreted at least one object with
greater total mass than the Milky Way disk ($m>5\times
10^{10}h^{-1}M_{\odot}$) over the redshifts $z\leq
3$ \citep{thin_disk_prob,disk_surv}.  It has been shown recently
\citep{disk_surv2, disk_surv3}, that the disk may reform after mergers
or even survive minor mergers, somewhat easing the problem of disk
survival.  Still, late-time bombardment of Milky-Way like haloes,
predicted within the \lcdm\ model, could still be considered as a concern
about the survival of thin-disk-dominated galaxies and the existence of
galaxies with unusually quiet merger history like our Galaxy.
Moreover, recent observations of the distant
Universe suggest that much of the stellar mass of bright galaxies was
already present in place at $z>1$, this presents the ``downsizing''
challenge for models of galaxy formation because massive haloes are
assembled late in the hierarchical clustering process intrinsic to the
CDM cosmology \citep{downsizing,downsizing_problem,
hierarchy_problem_star_mass_z_gt_1}. \sout{Finally} In addition there were also reports on
massive superclusters found at
intermediate redshifts ($z\sim1$), which seem to be very rare and extreme objects within standard
\lcdm\ model \citep{2sFGRS_superclusters_problem, 2dFGRS_final_data,
UKIDSS_z09_supercluster}. Finally there is a recent claim of possible observational evidence for a ``fifth force'' in DM sector \citep{lee}.

Even though we would not go as far as to call this a ``galaxy formation
crisis'',
there are noticeable tensions between what observations suggest
and what (cosmological simulations of) the \lcdm\ model seems to
predict. The ReBEL (\textit{daRk Breaking Equivalence principLe})
model discussed in recent years in the literature
\citep{GP1,GP2,PeeblesIDMDE,Farrar2007,rebel_nature} has the potential to alleviate
these tensions and to help the theoretical \lcdm\ model to better match
the observations. The ReBEL model is a slight modification of the
standard cold dark matter model. It introduces an additional long-range
scalar mediated force only in the dark sector. This has been studied in
numerical simulations \citep{NGP,LRSI1,LRSI2,Rebel,LRSI3} and these studies
have shown that the model can match observational properties of the
large-scale structures (like the one-dimensional Ly-$\alpha$ power
spectrum and the power spectrum of the SDSS\footnote{Sloan Digital Sky Survey - \url{http://www.sdss.org/}}
galaxies) as well as the standard \lcdm\ does, providing at the
same time interesting features with respects to the formation of galactic
structures. We also acknowledge that recently other authors performed
structure formation tests in similar scalar-interacting DM
scenarios \citep{Coupled_DE1,Coupled_DE2,Coupled_DE3,Coupled_DE4}, but they focused on models
containing scalar-field-like Dark Energy interacting with DM. Elementary
considerations related to the ReBEL model and the reionisation mechanism
showed that scalar-interacting DM could accommodate early reionisation of
the intergalactic medium \citep{reion1}. The low-resolution simulations
presented in the ReBEL literature have indicated that this model can
foster a picture in which structure formation starts at higher
redshifts. In this letter we will show that this is indeed the case. We
will present, for the first time in the literature, results of analysis
of the high-resolution N-body simulations of hierarchical DM halo
formation in the ReBEL framework.

\section{Self-interacting DM: The ReBEL model}

The theoretical ground of the ReBEL model was formed in two papers by
\citet{GP1,GP2}. We follow the phenomenological description of this
scalar-interacting DM model presented in \citet{NGP} and \citet{LRSI1}. We
consider a modified effective potential and
force law between DM particles of the forms
\beqa
\label{eqn:rebel-grav-mod1}\Phi(r) &=& -{G m\over r}\left(1 + \beta e^{-r/r_s}\right)\,,\\
\label{eqn:rebel-grav-mod2}F_{DM} &=& -{G m^2\over r^2}\left[1+\beta\left(1+{r\over r_s}\right)e^{-r/r_s}\right]\,,
\eeqa
where the two free parameters of the model are: $\beta$, the measure of
the strength of the scalar force compared to usual Newtonian gravity, and
$r_s$, the screening length of the additional force that is
\emph{constant} in a comoving frame. Previous considerations in the
literature of the subject provide a crude estimate of the values of this
free parameters to be of orders $\beta\sim 1$ and $r_s\sim 1\hmpc$. In
this short report we explore two possible values of the $\beta = 0.5,1$
and two values of the screening length parameter $r_s = 0.5,1\hmpc$. We
label the corresponding simulations runs as \emph{\lcdm, b05rs05,
b05rs1, b1rs1}, where b stands for $\beta$ and rs marks the $r_s$
parameter. In the \emph{\lcdm} run we use
$\beta=0$ (i.e. no scalar forces).  Equations
(\ref{eqn:rebel-grav-mod1},\ref{eqn:rebel-grav-mod2}) yield a simple
phenomenological description of the ReBEL addition to the CDM model.

\section{Numerical modelling}
\label{sec:numerics}

To follow the formation of structures within the ReBEL framework we use
an adapted version of the \verb#GADGET2# code \citep{Gadget2}. For the
detailed descriptions of the modifications made to the code we refer the
reader to our previous paper on this subject \citep{LRSI1}. We have
conducted a series of high-resolutions DM only $N$-body simulations
containing $512^3$ particles within periodic box of $32\hmpc$ comoving
width. The cosmology used in the simulations was the canonical \lcdm\
with $\Omega_m=0.3$, $\Omega_{\Lambda}=0.7$, $\sigma_8=0.8$ and $h=0.7$.
Thus our mass resolution is $m_p\simeq2.033\times
10^{7}h^{-1}M_{\odot}$. The force softening parameter was set to be
$\varepsilon=6\hkpc$. All simulations were started with the same initial
conditions at $z_{ini}=50$, and varied only in the ReBEL parameters. For
each run we saved 30 snapshot equally spaced in the logarithmic time
scale starting from $z=6.092$ to $z=0$. 

We used the MPI+OpenMP hybrid \verb#AMIGA# halo finder (\verb#AHF#) to
identify haloes and subhaloes in our simulation\footnote{AHF is freely
available from \url{http://popia.ft.uam.es/AMIGA}}. \texttt{AHF}
is the successor of the 
\texttt{MHF} halo finder by \citet{MHF}, a detailed description
of AHF is given in the code paper \citep{AHF}.  We note that
we have adjusted the code slightly to take the modified gravity of the
ReBEL models into account.  Starting with the halo catalogues at $z=0$
we construct the merger trees for each simulation by cross-correlating
the particles constituting the haloes in consecutive time steps.  For
each halo we record all progenitors and select the main progenitor as
the halo that maximises the merit function $N_1N_2/N_s^2$, where
$N_{1,2}$ and $N_s$ are the number of particles of the haloes and the
number of shared particles, respectively.

\section{The results}

In this \textit{Letter} we focus primarily on the mass accretion
histories of dark matter haloes and their masses, respectively. A more
in-depth study of the effects of the ReBEL model on their internal
properties will be presented in a companion paper (Hellwing, Knollman \& Knebe, in preparation).

\subsection{Mass accretion histories}

Following the scheme presented in \citet{Wechsler2002} (hereafter W02),
we fit each halo's mass accretion history (MAH) to the exponential law
of the form
\beq
\label{eqn:MAH_exp}
\tilde{M}(a) = \exp[\alpha(1-a^{-1})]\,,\quad \tilde{M}(a)\equiv {M(a)\over M_0}\,,
\eeq
where $\tilde{M}$ is the halo mass for a given cosmic scale factor $a$
expressed in terms of its final mass $M_0$ at $z=0$ and $\alpha$ is a
free parameter to be determined via fitting. 

The fit parameter can be used to express a characteristic redshift of
formation: \beq
\label{eqn:z_f}
z_* = {S\over\alpha} - 1\,,
\eeq
defined as the time when the logarithmic slope of the accretion rate
d$\log M/$d$\log a$ falls below some specified value $S$. In our studies
we adopt vale $S=2$ advocated in W02. Therefore our $z_*$ is the
redshift indicating the time when a dark matter halo has entered the
steady accretion phase of its mass accumulation (i.e.~the last major
merger should have taken place at $z>z_*$). The formation redshift ---
usually defined as the time when an object constituted
half of its present-day mass \citep[e.g.][]{Lacey.Cole.93} --- can now be
expressed in terms of $z_*$
\beq
\label{eqn:z_1/2}
z_{1/2} = {-\log{0.5}\over S}(z_*+1) = {-\log{0.5}\over\alpha}\,. 
\eeq
We will use $z_*$ and $z_{1/2}$ as obtained by fitting our
numerical MAHs to eqn.~\ref{eqn:MAH_exp} as estimators of the
steady accretion transitions and the formation times of our
halos. We acknowledge that our derivation of both these
redshifts is obviously not independent. To this extent, we also
derived the formation redshift $z_{1/2}$ by solely using the
numerical MAH of each halo finding the point where the mass drops
below $M_0/2$; the cross-correlation between these two $z_{1/2}$
values clearly follows a straight 1:1 relation with a scatter of less than
5\%. We therefore decided to stick to the value obtained by
eqn.~(\ref{eqn:z_1/2}).

In general, eqn.~(\ref{eqn:MAH_exp}) was shown to be a universal good
fit to the halos MAH's (e.g. W02), however \citet{Tasitsiomi_clusters}
pointed out that the simple exponential one parameter form of W02 is
not a good fit to the MAH of cluster haloes as well as haloes that
experienced recent major mergers. To not be biased by including such
objects in our study we removed all haloes whose $\chi^2$-value
deviated from the mean by more than 1$\sigma$ and all haloes whose Most Massive Progenitor (MMP) was not present in at least 24 (out of 30) snapshots, thus eliminating of order
$\sim 25$ per cent of all haloes in the original catalogues. 
We further limit the analysis to objects containing in excess of 500
particles (at redshift $z=0$) corresponding to a lower mass cut of
$\sim10^{10}h^{-1}M_{\odot}$. This leave us with $\sim 7000$
objects in the \emph{\lcdm} sample, and $\sim 4700$ objects in the \emph{b1rs1}
ReBEL run.

\begin{figure}
  \includegraphics[width=60mm,angle=-90]{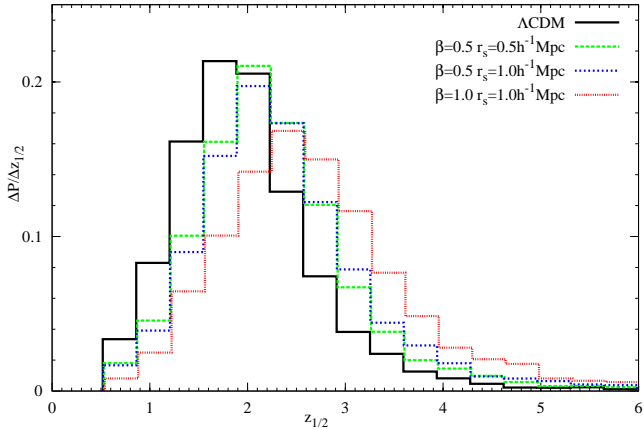}
  \caption{The probability distribution of
the formation redshift $z_{1/2}$ for all haloes containing at least 500 particles ($M_0>10^{10}h^{-1}M_{\odot}$). }
\label{fig2}
\end{figure}

\begin{table}
\caption{The mean and the $1\sigma$ scatter of the $z_f$ and $z_*$
distributions.}
\label{tab:z_mean}
\begin{center}
\begin{tabular}{lcccc}
\hline
The model ($\beta/r_s$) & $\langle z_{1/2}\rangle$ & $\sigma_{z_{1/2}}$ & $\langle z_*\rangle$ & $\sigma_{z_*}$\\
\hline
\lcdm & 1.83 & 0.8 & 4.32 & 2.3\\
$0.5/0.5\hmpc$ & 2.12 & 0.9 & 5.16 & 2.51\\
$0.5/1.0\hmpc$ & 2.21 & 0.9 & 5.4 & 2.7\\
$1.0/1.0\hmpc$ & 2.54 & 1.0 & 6.39 & 3.0\\
\hline
\end{tabular}
\end{center}
\end{table}

\begin{figure}
  \includegraphics[width=60mm,angle=-90]{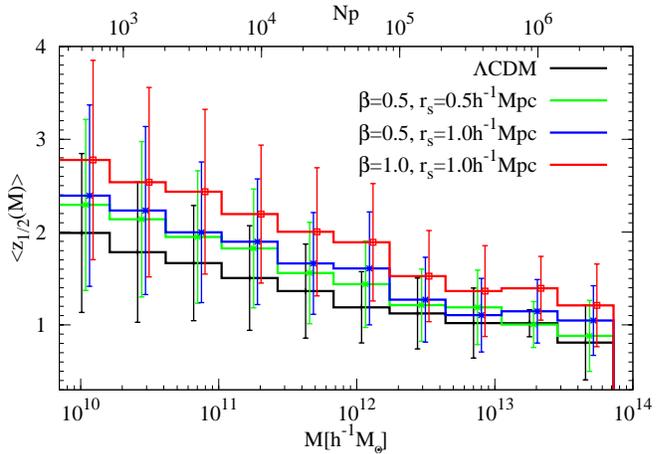}
  \caption{Hierarchical structure formation for all models. The relation
between the formation redshift ($z_{1/2}$) of objects and their masses
$M$. The error-bars represents $1\sigma$ scatter within the binned
distributions. The error-bars of the ReBEL model where slightly shifted
to the right for clarity.}
\label{fig:hsf}
\end{figure}

\subsection{Formation and Characteristic Redshifts} \label{sec:redshifts}

When plotting the distribution function of formation times
$z_{1/2}$ in Fig.~\ref{fig2} we observe a clear trend for the
mean of $z_{1/2}$ to shift to higher redshifts in the ReBEL
models. We like to note in passing that for the \lcdm\ model our
results agree with the ones presented in \citet[][their Fig.~2 and
3]{Lin2003}, when using the same mass bins as in said work.  

We calculated the same kind of distribution for the characteristic redshift $z_*$ (though not explicitly shown here) and list for both distributions the respective means and variances in Tab.~\ref{tab:z_mean}.
We note that for the strongest ReBEL model considered in this
\textit{Letter} the
mean value of $z_{1/2}$ is 39 per cent larger then the corresponding
$\lcdm$ value. This effect is
 even more pronounced for the
mean values of the $z_*$ distributions, the corresponding shift is more
than 48 per cent there.  The corresponding $1\sigma$ scatters are also
larger.
This is not surprising as
these two quantities are linked with linear relation of
eqn.~(\ref{eqn:z_1/2}). 
We also note that the magnitude of changes of $\langle
z_{1/2}\rangle$ and $\langle z_*\rangle$ in models \textit{b05rs05} and
\textit{b05rs1} are of comparable order. 
The reason for this is the following: our
sample is dominated by objects with masses
$M_0<10^{12}h^{-1}M_{\odot}$ whose virial radii are $r_{vir}<200\hkpc$;
therefore for majority of our objects the screening length of the ReBEL
force is much greater than their radii, hence for this class of objects
only the $\beta$ parameter measures overall enhancement of
scalar-attraction. 
To conclude this paragraph we denote that for the
parameter space probed by our simulations the $\beta$ parameter plays a
more important role on the examined quantities than the screening
length.

Viewing the shifts for formation and characteristic redshift in combination we determine that, on
average, haloes in the ReBEL model form earlier, thus arriving at
moderate and low redshifts with higher masses. Moreover haloes
with exceptionally long quiet accretion epochs are much more abundant
in the populations of our ReBEL models. We will return to the issue of larger masses below in Sec.~\ref{sec:halomassfunction}
  where we study the halo mass function at various redshifts.

\begin{figure}
  \includegraphics[width=57mm,angle=-90]{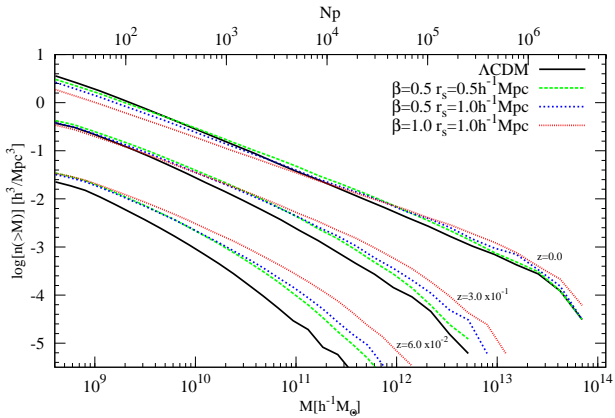}
  \caption{The cumulative mass functions for all haloes containing
    20 and more particles calculated at
    redshifts $z=6, 3$ and 0. Mass functions for redshifts
$z=6$ ($z=3$) were scaled down by $10^{-2}$ $(10^{-1})$
for clarity.}
\label{fig:cmfs}
\end{figure}

To check whether the hierarchical structure formation scenario is
preserved in the ReBEL models and how the effect of the earlier
structure formation depends on the halo mass we show in Fig.~\ref{fig:hsf} the
average $z_{1/2}$ binned in halo mass. We see that the
hierarchical character of the structure formation is preserved in all
the simulations runs, i.e.\ small mass objects form first with higher
mass objects forming later. In all the ReBEL runs the formation
redshifts are higher for all considered masses, but the effect is
strongest for less massive haloes. However, we also notice a clear
dependence on the parameters of the ReBEL model. For small $\beta$
and small $r_s$ the effect of shifting structure formation
to the higher redshifts is less pronounced. This is of course 
expected, since the screening length acts here as the effective radius
of an enhanced accretion and the $\beta$ measure the strength of this
enhancement. It is interesting to note, that for the \textit{b05rs05,
b05rs1} runs the aforementioned effect disappears at some fixed mass
scale. In other words, we see that the averaged ReBEL $z_{1/2}$ values start
to agree with the \lcdm\ values starting from mass $M\sim 5\times
10^{12}h^{-1}M_{\odot}$. It is only the \textit{b1rs1} model that
deviates throughout the whole mass range probed by our simulations.
Nevertheless we observe that in the \textit{b05rs05, b05rs1} simulations
the haloes with masses $10^{11}h^{-1}M_{\odot}<M<5\times
10^{12}h^{-1}M_{\odot}$ form at significantly higher redshift compared
to the standard CDM model. For the \textit{b1rs1} case even haloes with
masses comparable to a mass of a group of galaxies
$10^{13}h^{-1}M_{\odot}<M<10^{14}h^{-1}M_{\odot}$ acquire most of their
mass at relatively earlier times, however we denote that at masses
$M>10^{13}h^{-1}M_{\odot}$ the signal is dominated by small-number
effects, i.e. we only have of order $\sim 30$ objects in that mass range: to more thoroughly check the
differences in \lcdm\ and ReBEL models at this mass scale simulations
with a larger box size are required.

\subsection{The Evolution of the Mass Function} \label{sec:halomassfunction}
As speculated before, haloes in the ReBEL models not only form
  earlier but also carry a larger mass. To actually confirm and
  quantify this previously made claim (cf. Sec.~\ref{sec:redshifts})
  we plot in Fig.~\ref{fig:cmfs} the cumulative mass functions (CMFs) at the
  redshifts $z=6, 3$, and $0$.  

We notice that for $z=6$ and $z=3$ all ReBEL's CMFs have higher
amplitudes than the fiducial \lcdm\ CMF in the whole probed mass range:
At high redshifts the ReBEL model produces more non-linear structures
compared to the standard CDM model. However structure formation in the
\lcdm\ paradigm seems to be 'catching' up with ReBEL at lower redshifts.
This can again be seen in Fig.~\ref{fig:cmfs}: the discrepancies of the
amplitudes at redshift $z=0$ are much suppressed in comparison to the
high redshift CMF's. In addition, we note an interesting feature at the
final redshift: while the CMFs have a higher amplitude in the ReBEL
models for redshifts $z=6$ and $3$ across the whole mass range, at $z=0$
only the high mass end of the ReBEL mass functions shows an excess
relative to the fiducial \lcdm\ model. The low mass end of the ReBEL
CMFs at this redshift show deficiency of haloes compared to the LCDM.
The latter effect indicates that in the ReBEL model some of the low
mass haloes were already used in structure formation processes as
bricks for more massive objects.  All this taken together clearly points
out that the process of rapid non-linear structure formation takes place
at later times in the \lcdm\ model compared to model with the ReBEL
forces.

\section{Discussion}

The main conclusion of this \textit{Letter} is that a slight
modification to the CDM model in the form of scalar interactions can
accommodate more gravitationally bound structures at high redshifts
(e.g. fig.~\ref{fig:cmfs}) which in turn promotes an early reionisation
of the Universe. This can be expressed by examining the $\eta$ factor
that describes the efficiency of reionisation of the Inter Galactic
Medium (IGM) \citep{reion1,Cen2003}:
\beq
\label{eqn:eta_factor}
\eta\equiv{c_{\star}f_{esc}(\textrm{d}f_M/\textrm{d}t)e_{UV}\over
  C(1+z)^3}\,,
\eeq
where $c_{\star}$ is the star formation factor efficiency, $f_{esc}$ is
the ionising photon escape fraction, d$f_M/$d$t$ is the halo formation
rate, $e_{UV}$ is the ionising photon production efficiency and $C$ is
the gas clumping factor. Here the numerator reflects the ionising photon
production rate while the denominator is the ionising photon destruction
rate. When $\eta$ exceeds a certain threshold, the IGM becomes ionised.
As the ReBEL models exhibit a richer structure at $z=6$, their formation
rate must have been higher than the one of the \emph{\lcdm} run.  This
would make it easier to reach the threshold of reionisation at high
reshifts in the ReBEL framework.

We further have to acknowledge that earlier formation of the DM haloes
fostered by the ReBEL model, results in shifting a
significant part of the violent merger events to higher redshifts. This
indicates that there are many more haloes with extended steady accretion
histories than expected in the standard \lcdm\ structure formation
scenarios. This would help thin-disk dominated galaxies to preserve
their disks at low redshifts and would also imply that the exceptionally
long period of peace experienced by the Milky-Way is no more so
extraordinary \citep{Milky-Way}.

Finally, we have also shown that long-range scalar interactions between
DM particles increase to some extent the masses of DM haloes. That is
also an important effect that may help to understand
the existence of objects like 
massive superclusters at moderate redshifts. The
simulations presented in this \textit{Letter} however do not encompass a
large enough volume to study the formation of these massive objects in
detail.

In conclusion, we have shown that the ReBEL model allows to alleviate
some of the tension between predictions of the \lcdm\ and observations
\citep{reion2,reion3,thin_disk_prob,disk_surv,2sFGRS_superclusters_problem},
however, as we only focused on the issue of formation time, it still
needs to be investigated if and how internal properties of dark matter
haloes are affected by the ReBEL formalism.  One interesting quantity is
the halo concentration parameter, as it correlated with the halo
formation time \citep[eg.][]{Wechsler2002}.   Preliminary work shows this
to indeed be the case, with the  mean 
concentration $\langle c_{vir}\rangle$ being $9$ ($11$, $11$, $13$) for \lcdm\ (\textit{b05rs05},
\textit{b05rs1}, \textit{b1rs1}) for haloes with masses $M\sim2\times 10^{12} h^{-1}M_{\odot}$,
respectively.  However the issue of the halo concentration parameters require deeper studies and, since a more detailed study of the haloes internal properties
is in progress, we will cover the concentration parameter issue in a forth-coming work.

\section*{Acknowledgements}

WAH would like to acknowledge the hospitality of the Departamento de
F\'isica Te\'orica at the Universidad Aut\'onoma de Madrid during his
short stay there. Some of the simulations used in this research was
performed on the \verb#halo# cluster at the Warsaw University
Interdisciplinary Center for Mathematical and computational Modelling.
WAH is supported by the Polish Ministry of Science grants no.~NN203
394234 and NN203 386037. SRK acknowledges support by the Ministerio de
Ciencia e Innovacion (MICINN) under the Consolider-Ingenio, SyeC project
CSD-2007-00050. AK is supported by the MICINN through the Ramon y Cajal
programme.

\bibliographystyle{mn2e}
\bibliography{lrsi_letter}

\bsp

\label{lastpage}

\end{document}